\documentclass[11pt]{article}
\usepackage{amsmath}
\usepackage{amsfonts}
\usepackage{amssymb}
\usepackage{graphicx}
\usepackage{subfigure}
\usepackage{graphicx}
\usepackage{dcolumn}
\usepackage{bm}

cm \headheight=0. mm

\def\be{\begin{equation}}
 \def\ee{\end{equation}}
\def\bea{\begin{eqnarray}}
\def\eea{\end{eqnarray}}

\def\l{\label}
\def\ov{\over}
\def\R{\rho}

\begin{document}
\begin{center}
\LARGE {QCD phase transition with a power law chameleon scalar field in the bulk}
\end{center}
\begin{center}
{\bf$^a$T. Golanbari\footnote{t.golanbari@gmail.com}},
{\bf $^{b}$A. Mohammadi\footnote{abolhassanm@gmail.com}},
{\bf $^{a} $Kh. Saaidi\footnote{ksaaidi@uok.ac.ir}},
\\
{\it$^a$Department of Physics, Faculty of Science, University of Kurdistan,  Sanandaj, Iran}\\
{\it $^b$Young Researchers and elites Club, Sanandaj Branch, Islamic Azad University, Sanandaj, Iran.}\\
\end{center}
\vskip 1cm
\begin{center}
{\bf{Abstract}}
\end{center}
\ \ \ In this work, a brane world model with a perfect fluid on brane and a scalar field on bulk has been used to study quark-hadron phase transition. The bulk scalar field has an interaction with brane matter. This interaction comes into non-conservation relation which describe an energy transfer between bulk and brane. Since quark-hadron transition curly depends on the form of evolution equations therefore modification of energy conservation equation and Friedmann equation comes into some interesting results about the time of transition. The evolution of physical quantities relevant to quantitative of early times namely energy density $\rho$ temperature $T$ and scale factor $a$ have been considered utilizing two formalisms as crossover formalism and first order phase transition formalism. The results show that the quark-hadron phase transition in occurred about nanosecond after big bang and the general behavior temperature is similar in both of two formalism. \\



\section{Introduction}
\ \ \ Recently, models of the Universe with extra dimensions have  attracted serious attention. Among the most well-known of these models are the original brane-world scenarios  considered by Randall and Sundrum in 1999 \cite{1, 2}.
These  scenarios present an interesting picture of the Universe in which the standard matter fields are confined to a  four-dimensional hypersurface (the brane) embedded in higher-dimensional space-time (the bulk). The graviton, by exception, is allowed to propagate in the bulk. In this model the four-dimensional Friedmann equation can be recovered in low energy limit with a positive tension for brane embedded in five-dimensional Anti-de Sitter space-time. On the other hand, the model suggests that the dynamic of the early Universe (high energy regime) should be modified related to the contribution of quadratic term of energy density and bulk Weyl tensor. This result is so important because we have some changes in the basic equations which describe the cosmological and also the astrophysical dynamics of the Universe, and this has been recently the subject of so many research works {\cite{3, 4, 5, 6, 7}}. \\
Drawing on motivations from particle physics, there has been some recents interest in the possibility that non-gravitational matter-energy can be exchanged between the brane and  bulk \;\cite{8, 9}. In this case, the  evolution of the matter fields on the brane is modified by the introduction of  new terms in the energy-momentum tensor which describes the matter-energy  exchange between the brane and the bulk.  In Ref.\;\cite{10}, it was found that such a  model can explain the present cosmic acceleration as a consequence of the flow of matter from the bulk to the brane. Interestingly, the model also seems to predict the observed suppression of the lowest multipole moments in the power spectrum of the cosmic microwave background. Some features of the brane-world with bulk-brane matter-energy transfer have already been  studied. For example, the cosmological evolution of
a brane with  a chameleon scalar field in the bulk was studied in Ref.\;\cite{11, 12},  the cosmological evolution of
a brane with a general matter content in the bulk was studied in Ref.\;\cite{10}, and  the reheating of a brane-world Universe with bulk-brane matter-energy transfer  was studied   in Refs.\;\cite{13}.\\
Another suggestion which leads to some interesting models even in standard model of cosmology is scalar field. The problem of scalar field in brane-world scenario attracted great interests {\cite{14, 15, 16}}.
According to string theory, it is also possible that the brane-world models contain bulk scalar field which is free to propagate in extra dimensions {\cite{17}}.  This scalar field also could interact with matter and tension of brane which is called dilaton scalar field {\cite{17, 18, 19}}. It is natural and attractor to study the effect bulk scalar field on the brane evolution and it is not unexpected that these kinds of models represent some interesting consequences. Before that it is remarkable to mention that there is such a scalar field in standard cosmology with non-minimal coupling to matter which called chameleon scalar field. Based on that, we called the bulk scalar field as a chameleon scalar field \cite{20, 21}. The model also comes to some absorbing results about the brane evolution. In \cite{22}, it is shown that although bulk is free of matter but due to the interaction to matter, the scalar field takes a mass which depends on the brane energy density. This fact comes to this results that the mass of scalar field on four-dimensional brane is not small and therefore correction to the Newton law cannot be large because of propagation of scalar field in the bulk. There can be some other appealing feature for this model in reheating area as well, refer to {\cite{22}}. In this area, the scalar field decay to other particles. The matter scalar field, with zero bare mass, will find an effective mass as $m_{\psi}=g\phi(t)$ which in this model $m_{\psi}$ is much smaller than $m_{\phi}$ and so according to {\cite{23}} this plays the crucial role in the reheating process. In addintion, the model is able to predict and accelerated expansion in late time and an exponential expansion in the early times which could be related to inflationary epoch of the universe. The interesting point about this kind of scalar field is that the conservation relation is modified (in addition to the Friedmann equation) and the model possesses this ability to suggest the energy-exchange term. Then, there is no need to recommend a manual term. The energy-exchange is directly resulted from the introduced term in the action which describes the interaction between matter and scalar field, and this can be one of the advantages of the model. Based on above mentioned feature for this model of scalar field in braneworld scenario, and its compatibility  with string theory suggestion it is found out interesting to study the ability of such a models in other era of universe evolution.

On the other hand,  according to the  standard model of cosmology,
as the  Universe expanded and cooled it passed through  a series of symmetry-breaking phase transitions which might have  generated  topological defects \;\cite{24}.   This early Universe  phase transition could have been of first, second, or higher order, and it  has been  studied in detail  for over three decades   \cite{25,26,27,28,29,30,31,32,33,34,35,36,37,38,39,40,41,42,43,44,45}.  We note that the possibility of no phase transition was considered  in Ref.\;\cite{46,47,48,49,50}. The phase transition using both first order formalism and crossover formalism has been considered in standard cosmology \cite{51,52,53,54} with viscosity effect {\cite{55,56,57}} that causes a kind of non-conservation relation. Furthermore, the first order and the crossover approaches to phase transitions were studied in the brane-world \cite{24, 58}, DGP brane-world \cite{59},  and in Brans-Dicke models of brane-world gravity \cite{60} without any bulk-brane energy transfer.

In the present paper,  we study the quark-hadron phase transition (or \emph{QCD phase transition}) in a brane-world with a chameleon scalar field in the bulk.  We find that, in this model, the non-minimal coupling between  the scalar field and matter modifies the energy  conservation equation for  the matter fields on the brane. 
This fact, together with the Friedmann equation in the brane-world scenario (which is different  from the  Friedmann equation in the   four-dimensional case), causes an increased expansion in early times and this  has an important effect on the cosmological phase transition.  The model is a generalized form of some other models {\cite{24,60,61}} which have considered phase transition and their works have been published in some valid journals such as \textit{Phys. Rev. D, Nucl. Phys. B, Class. Quant. Grav.}. Moreover, the following reasons lead one to this result that our model possesses this suitability to be considered
\begin{itemize}
  \item The model is motivated from extra dimensions models which have a special place in the physicist research works.

  \item The model includes a bulk scalar field based on the suggestion of string/M theory.

  \item The presence of bulk-brane energy-exchange term in the model is based on the prediction of particle physics.

  \item Avoiding freely selection of energy-exchange term.
\end{itemize}
This paper is organized as follows. In Sec.\;2, we  introduce the model and derive the equations of motion. Sec.\;2 is related to the study of the QCD phase transition (smooth crossover  approach), and in Sec.\;3.  We review the first-order phase transition and consider it in  our model in Sec.\;5 we summarize our results.


\section{General Framework}
\ \ \ In accordance with the brane-world scenario, we treat our usual four-dimensional spacetime (the brane) as an embedded submanifold  of  a five-dimensional spacetime (the bulk). In order to accommodate the chameleon scalar field $\phi$ in the bulk, we postulate the following action
\begin{equation}\label{1a}
A=\int d^5x \sqrt{-g} \left\{\frac{1}{2}R^{(5)}-\Lambda_5 - \frac{1}{2}(\nabla\phi)^2-V(\phi)\right\} - \int d^4x \sqrt{-h} \mathcal{L}_4 .
\end{equation}
We have expressed the action in terms of the five-dimensional coordinates $(x^0,x^1,x^2,x^3,y)$ on the bulk, such that the brane is the hypersurface described by   $y=0$. Moreover, we have assumed units in which the five-dimensional Planck mass equals one. The signature of the five-dimensional metric $g_{\mu\nu}$ is  $-++++$. The first term corresponds to the Einstein-Hilbert action generalized to five dimensions together with a scalar field $\phi$ (the chameleon scalar field) whose potential is given by the function $V(\phi)$, which is assumed to be almost flat. Note that, due to the dimensional units of the energy-momentum tensor, the scalar field no longer has dimension $M$, but rather $M^2$.
Also, the dimension of the potential is $M^5$. The  determinant of $g_{\mu\nu}$  and the five-dimensional Ricci scalar are denoted by $g$ and $R^{(5)}$ respectively. $\Lambda_5$, in the first integral, is the bulk cosmological constant. The second term describes the coupling between the scalar field  and the matter fields. The metric $g_{\mu\nu}$ in the bulk  induces a metric $h_{\mu\nu}$ on the embedded four-dimensional brane. The determinant of $h_{\mu\nu}$ is denoted as $h$.
$\mathcal{L}_4$, in the brane action, is defined as $\mathcal{L}_4=L_m(\psi_m,\tilde{h}_{\mu\nu}) - \lambda$ where the matter field is $\psi_m$ and the matter  Lagrangian on the brane is denoted by  $L_m$, and $\lambda$ denotes brane tension. We assume that the non-minimal coupling between the  scalar field and the matter field is given by
\begin{eqnarray}\label{2a}
\tilde{h}_{\mu\nu}&:=&S(\phi)h_{\mu\nu},
\end{eqnarray}
where $S(\phi)$ is an analytical function of the scalar field.

By varying the action (\ref{1a}) with respect to $g_{\mu\nu}$, one obtains the five-dimensional Einstein field equation
\begin{eqnarray}\label{3a}
\phantom{|}^5G_{\mu\nu}&=&T_{\mu\nu}^{(\Lambda)}+T_{\mu\nu}^{(\phi)}+T_{\mu\nu}^{(b)},
\end{eqnarray}
where $T_{\mu\nu}^{\Lambda}$ and $T^{(\phi)}_{\mu\nu}$ respectively denote the bulk cosmological constant energy-momentum tensor and the total energy-momentum tensor for the scalar field, which is defined as
\begin{equation}
T^{(\Lambda)\mu}_{\phantom{(\Lambda)\mu}\nu}= (-\Lambda_5,\Lambda_5,\Lambda_5,\Lambda_5,\Lambda_5)
\end{equation}
and
\begin{eqnarray}\label{4a}
T^{(\phi)}_{\mu\nu}&=&\nabla_\mu \phi \nabla_\nu \phi - g_{\mu\nu}\left\{\frac{1}{2}(\nabla \phi)^2 + V(\phi)\right\}.
\end{eqnarray}
 $T_{\mu\nu}^{(b)}$ is  the brane energy-momentum tensor which is given by
\be\l{5a}
T^{(b)}_{\mu\nu}= \frac{2}{\sqrt{-h}} \frac{\partial \left( \sqrt{-h} \mathcal{L}_4 \right)}{\partial h^{\mu\nu}} \delta(y).
\ee
In fact, one can write
\be\l{6a}
T^{(b)\mu}_{\phantom{(b)\mu}\nu}= \delta(y) {\rm diag} (-\rho_b,p_b,p_b,p_b,0).
\ee
Here, $\rho_b$ and $p_b$ denote  the brane energy density
and pressure respectively. These quantities include the tension $\lambda$ of the  brane, so that
\begin{eqnarray}\label{7}
\rho_b&=&\rho_m+\lambda \nonumber \\
p_b&=&p_m-\lambda,
\end{eqnarray}
where $\rho_m$ and $p_m$ denote the  matter density and pressure respectively; see Refs.\;{\cite{62,63,64}}.
Varying the action (\ref{1a}) with respect to $\phi$ leads to the  field equation for the scalar field
\begin{eqnarray}\label{7a}
\phantom{|}^5\Box\phi&=&V_{,\phi} +{1\over 2}{\tilde{S}(\phi)\over S}T^{(b)} \delta(y),
\end{eqnarray}
where $\phantom{|}^5\Box$ is the five-dimensional d'Alembertian, $\tilde{S}(\phi) := dS(\phi)/d\phi$ and $T^{(b)}$ is the trace of brane energy momentum tensor.

 We introduce a Friedmann-Lema$\hat{\rm i}$tre-Robertson-Walker (FLRW) metric with a maximally symmetric 3-geometry $\gamma_{ij}$
\begin{equation}\label {8a}
ds^2=-n^2(t,y)dt^2+a^2(t,y)\gamma_{ij}dx^idx^j+dy^2.
\end{equation}
Our brane is the hypersurface  given by $y=0$,  and  we also  take
$Z_2$ symmetry  into  account. It should be mentioned that the metric coefficients are
continuous but their first derivatives with respect to $y$ are
discontinuous, and  furthermore their second derivatives with respect to $y$
include the Dirac delta function. Substituting the above metric into the field equations,
one gets that the non-vanishing components of Einstein tensor are
\begin{eqnarray}\l{9a}
^{(5)}G_{00}&=&-\frac{3}{a^2}\Big\{ -\dot{a}^2+a'^2n^2+a a'' n^2
\Big\},\\
^{(5)}G_{ij}&=&\frac{\delta_{ij}}{n^3} \Big\{
2a\ddot{a}n+2\dot{n}a\dot{a}+2n^2n'aa' - n\dot{a}^2+a'^2n^3+2a a'' n^3+n^2a^2 n'' \Big\} ,\l{10}\\
^{(5)}G_{05}&=&-\frac{3}{an}\Big\{ \dot{a}'n-n'\dot{a} \Big\},\l{11}\\
^{(5)}G_{55}&=&\frac{3}{a^2n^3}\Big\{
-a\ddot{a}n+\dot{n}a\dot{a}+n^2n'aa'-n\dot{a}^2 + a'^2n^3 \Big\}.\l{12}
\end{eqnarray}
Here, dots denote derivatives with respect to  time and primes denote
derivatives with respect to the fifth coordinate. Since the second
derivative of the metric involves the Dirac delta function, we write, following Ref. {\cite{62,63}}
\be\l{13}
a''=\hat{a}''+[a']\delta(y),
\ee
 where $\hat{a}''$ is the
non-distributional part of the double derivative of $a(t,y)$, and
$[a']$ is the jump in the first derivative across $y=0$, which can be expressed schematically as
\be\l{14}
[a']=a'(0^+)-a'(0^-).
\ee
 The junction functions can be obtained
by matching the Dirac delta functions in the components of Einstein
tensor with the components of the brane energy-momentum tensor. From
the $(0,0)$ and $(i,j)$ components of field equations, we have,
respectively
\begin{eqnarray}\label {15}
\frac{[a']}{a_0}&=&-\frac{1}{3}\rho_b,\\
\frac{[n']}{n_0}&=&\frac{1}{3}(2\rho_b+3p_b),\l{16}
\end{eqnarray}
where  the
subscript $``0"$ means that we evaluate on   $y=0$. These equations are the same as  the junction
relations in Ref.\;{\cite{62,63}}. On the other hand, according to Eq.\;(\ref{7a}), we get
\begin{equation}\label {17}
\frac{\ddot{\phi}}{n^2}-\phi'' + \left(
\frac{3\dot{a}}{an^2}-\frac{\dot{n}}{n^3}
\right)\dot{\phi}-\left( \frac{n'}{n}+\frac{3a'}{a} \right)\phi'=
 - V_{,\phi}(\phi)-{1\over 2}{\tilde{S}(\phi)\over S}T^{(b)} \delta(y).
\end{equation}
Matching the Dirac delta functions on both sides of Eq.\;(\ref{17}) gives,
for $y=0$:
\begin{equation}\label {18}
[\phi'_0]= {1\over 2}{\tilde{S}(\phi_0)\over {S(\phi_0)}}T^{(b)}.
\end{equation}
 Also, because we have  $Z_2$ symmetry, one can
obtain $a'_0$, $n'_0$ and $\phi'_0$ functions from the junction
conditions.

The $(0,5)$ component of the Einstein tensor reads:   $$^{5}G_{05}= T^{(\phi)}_{05}=\dot{\phi} \phi',$$ ($T^{(b)05}=0$). Now, by substituting $\dot{a}'$ and $n'$ from   (\ref{15}) and (\ref{16}), and using  $Z_{2}$ symmetry, one can obtain  generalized continuity equation on the brane
\begin{equation}\label{19}
\dot{\rho_b}+3H(\rho_b+p_b)=2\dot{\phi_0}\phi'_0.
\end{equation}
This equation is the modified continuity relation of energy density and indicates a bulk-brane energy transfer.
   Note that we have set $n_0 =1$  in all  relations of
 this section.

From  the $(5,5)$ component of the Einstein tensor, one gets the second order (or generalized) Friedmann equation
\begin{equation}\label{20}
\frac{\ddot{a}_0}{a_0}+\frac{\dot{a}_0^2}{a_0^2}=-\frac{1}{36}\rho_b(\rho_b+3p_b)
- \frac{\Lambda_5}{3} - \frac{1}{3}\bigg [\frac{\dot{\phi}^2_0}{2}+
\frac{\phi'^2_0}{2} - V(\phi_0)\bigg].
\end{equation}
 The first Friedmann equation on the brane is obtained
from the  $(0,0)$ component of the field equation
\begin{equation}\label{21}
H^2=\left(\frac{\dot{a_0}}{a_0}\right)^2=\frac{1}{36}\rho_b^2+ \frac{\Lambda_5}{3} +
\frac{1}{3}{\Bigr [}\frac{\dot{\phi}^2_0}{2}+ \frac{\phi'^2_0}{2}
+ V(\phi_0){\Bigl ]}.
\end{equation}
Here we have assumed that  the non-distributional part of the  double derivative
of $a(t,y)$ with respect to the fifth coordinate, $\hat{a}''_0$, is zero.
It is seen that equation (\ref{21})  agrees with the results obtained in Refs.\;\cite{65} and \cite{66} for brane world cosmology and is completely different from the  standard cosmological model because in standard cosmology, $H \propto\sqrt{\rho}$   rather than $ {\rho}$. It is seen that in Eqs.\;(\ref{17}), (\ref{19}) and (\ref{21}) there is an additional  free parameter.

To determine behavior of scalar field, one should solve scalar equation (\ref{17}). A kind of interaction between matter and scalar field which has been assumed in the model brings scalar field dependence for brane matter density  {\cite{11,12}}.  Hence finding out the exact solution of equation encounter difficulties. Therefore in these cases and to simplify the problem, a specific function for scalar field is supposed.
 In accordance with Ref.\;\cite{67,68}, we shall assume that the  scalar field can be expressed  as a power  of the scale factor
\begin{equation}\l{22}
\phi_0 =N_0\Big[{a(t) \ov a(t_0)}\Big]^{\xi} =  N a^{\xi}.
\end{equation}
Here, $a(t_0)$ is the scale factor at the present time, $a(t_0) =1$, $N$, and $\xi$ are  constant.
Using Eqs.\;(\ref{7}), (\ref{18}), and (\ref{22}) we can rewrite Eqs.\;(\ref{21}), (\ref{17}) and (\ref{19})   as (by imposing Randall-Sundrum fine-tuning, namely $\Lambda_4=\frac{1}{12} \lambda^2  + \Lambda_5=0$)
\begin{eqnarray}\label{23}
H&=&{1 \ov \sqrt{6(6 -(\xi\phi_0)^2) }}{\Bigg [}  \rho_m \Big( 2\lambda+ {\rho_m} \Big) + {3 \ov 8}{\Big (}{\tilde{S}(\phi_0) \ov S(\phi_0)}{\Big)}^2 {T^{(b)}}^2 +12 V(\phi_0){\Bigg ]}^{{1}/{2}}\\
\ddot{\phi}_0 &+& 3H\dot{\phi}_0 =   {1 \ov 24}\frac{\tilde{S}(\phi_0)}{S(\phi_0)}  T^{(b)2} -V_{,\phi}(\phi_0),\l{24} \\
\dot{\rho}_b &+& 3H(\rho_b + p_b) =  { \xi H T^{(b)} \ov  2}\Big[ {\tilde{S}(\phi_0) \ov S(\phi_0)} \phi_0\Big].\l{25}
\end{eqnarray}
In order to gain better insight, we choose  $S(\phi_0) = s_0\phi^m$. For this choice,  Eqs.\;(\ref{23}), (\ref{24}) and (\ref{25}) are rewritten as
\begin{eqnarray}\label{26}
H&=&{1 \ov \sqrt{6(6 -(\xi\phi_0)^2) }}{\Bigg [}  \rho_m \Big( 2\lambda+ {\rho_m} \Big) + {3 \ov 8}{\Big(}{m T^{(b)}\ov \phi_0}{\Big)}^2  +12 V(\phi_0){\Bigg ]}^{{1}/{2}}\\
\ddot{\phi}_0 &+& 3H\dot{\phi}_0 =   {1 \ov 24}\frac{m}{\phi_0}  T^{(b)2} -V_{,\phi}(\phi_0),\l{27}\\
\dot{\rho}_b &+& 3H(\rho_b + p_b) ={m \ov 2}\xi H  T^{(b)} .\l{28}
\end{eqnarray}
A specific function for scalar field potential density  should be assumed to  continue our study in subsequent sections. Two  scalar field  potential energy densities, exponential and inverse power law, are commonly  used in discussion of  the chameleon mechanism. Here we consider  the inverse  power law potential energy density \cite{69,70}
\begin{equation}\label {42}
V(\Phi)=M^5 \left( \frac{{M^2}}{\phi}\right)^{\alpha},
\end{equation}
where $\alpha>0$, $M$ is a constant  mass scale and  the scalar field, $\phi$, has $M^2$  dimension. The authors of Refs.\;\cite{20,21,71} consider the solar system constraints  for a model  with this potential and find out that for small values of $\alpha  \in (0 ,  2)$  the magnitude of $M$ is  $\sim 10^{-3}$ eV. Therefore, the potential may be written as
\begin{equation}\l{43}
V(\phi) \sim {10^{-3(2\alpha +5)} \over \phi^{\alpha}}   \;\ {\rm eV}^5 .
\end{equation}
 Since the characteristic energy density scales of   other quantities such as  $\rho$, $p$, $\lambda$, and  constants of the model,    are of order a MeV,  the scalar field potential energy density term is very small compared  to other terms in the Lagrangian density and we can  ignore  it.

\section{QCD Phase Transition}\label{III}
\ \ \ In this section, we are going to examine   the physical quantities related to the quark-hadron phase transition. The results  will be applied in the context of a brane-world scenario with a chameleon scalar field in the bulk.  The phase transition in QCD can be characterized by the  truly singular behavior of the partition function, and may be either   a first or second order phase transition. It can only be  a crossover with rapid changes in  observable that  strongly depend  on the values of the quark masses.

To study the quark-hadron phase transition, we need equation of state for  matter   in both quark and hadron phase regimes. There exist various procedures  for finding equations  of state (EoS). In most of the recent calculations for $2+1$ flavor QCD, EoS has  been estimated.   The most extensive calculations of EoS have been performed with  fermion formulation
on lattice with temporal extent  $N_t = 4, 6$ \cite{72,73,74,75},
 $N_t=8$ \cite{76} and $N_t= 6, 8, 10$ \cite{77}. In the high temperature region, $T > 250$ MeV, the trace anomaly can be  calculated accurately, so one can use the lattice data for the trace anomaly in this region in order  to construct a realistic equation of state. In the low temperature region, the trace anomaly is affected by  a large discretization effect, but the  hadronic resonance gas (HRG) model is well-suited for  building a  realistic equation of state  in the low temperature  region \cite{78}.\\
Since in brane world scenario, all standard particles are confined on a four-dimensional hypersurface, then equation of state for matter in brane  world scenario  is the same as equation of state  in the standard four-dimensional model of cosmology. In fact the equation of state comes from the matter energy-momentum tensor, and presence of extra dimension in the model has no effects on this tensor. There are some other works which the authors consider quark-hadron phase transition in brane world scenario \cite{24, 58, 60} and they have applied the same equation of state in the four dimensional  in their works. \\
\subsection{High temperature region}
As noted above, one can use the lattice data for the trace anomaly in order to find  the  equation of state in the high temperature region, $T > 250$ MeV \cite{78}. In this regime, radiation-like behavior can be seen, and   the data  can be fit to a simple equation of state which reads as follows
\bea\l{29}
\R(T)&\thickapprox & \alpha_{r}T^4,\\
p(T) &\thickapprox & \sigma_{r}T^4,\l{30}
\eea
where $\alpha_{r} = 14.9702\pm 009997$ and $\sigma_{r} = 4.99115\pm 004474$.\\
Substituting Eqs.\;(\ref{29}) and (\ref{30}) into (\ref{28}) we obtain
\begin{equation}\l{31}
H=-{ 4 \alpha_r T^3\dot{T}\ov A_0T^4 +A_1},
\end{equation}
where
$$A_0 = 3(\alpha_r + \sigma_r) - \frac{m}{2}\xi (3\sigma_r - \alpha_r)$$ $$ { A_1= 2m\xi\lambda}.$$
By integrating  (\ref{31}) we have
\begin{equation}\l{32}
a(T)= a_c \Big( A_0 T^4 + A_1 \Big)^{-{\alpha_r/A_0}},
\end{equation}
where $a_c$  is a constant of integration, and from the modified Friedmann equation (\ref{26}), and (\ref{31})  the time derivative of temperature, $\dot{T}$, is
\begin{equation}\l{33}
\dot{T} = -\frac{A_0T^4 +A_1}{4\alpha_r T^3 \sqrt{6\big(6-(\xi \phi)^2}\big)}
    \Bigg\{  \alpha_r T^4 \Big( 2\lambda+ \alpha_r T^4 \Big)
   +  \frac{3 m^2}{8\phi^2} \Big[(3\sigma_r - \alpha_r)T^4-4\lambda\Big]^2\Bigg\}^{1/2}
\end{equation}
this  expression describe the behavior of temperature as a function of time in  the  brane world universe with
chameleon scale field in the bulk in the quark phase. The transition region in the crossover regime can be
defined as the temperature interval 250 MeV $ < T <$ 700 MeV.
\begin{figure}[ht]\label{0}
\centerline{ \includegraphics[width=7cm] {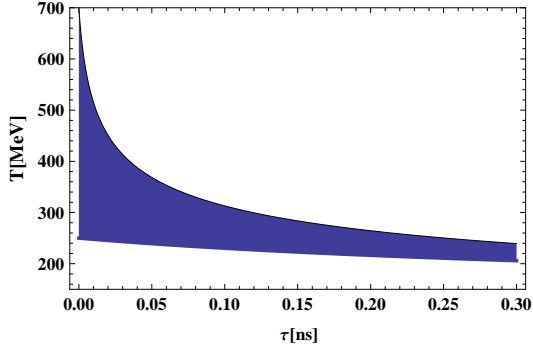}}
\caption{ $T$ versus $\tau$ according to high temperature region of the smooth crossover procedure in the brane gravity with chameleon scalar field in the bulk for $\lambda =10^9 $ MeV$^4$.  We have set $N =0.2\times 10^3 $, $\xi = 0.0002 $, and $m=1$. }
\end{figure}
We have plotted the numerical results of Eq.\;(\ref{33}) in Fig.\;1 for $\lambda =10^9 $ MeV$^4$ in the interval 250 MeV $ < T <$ 700 MeV. The plot shows two curve for Eq.(\ref{33}) related to two initial value of temperature as $T=700$ MeV and $T=250$Mev, and the interval space between this two curves is shaded. This  figure shows  the effective temperature of the Universe as a function of cosmic time in the quark-gluon phase (QGP)  according to  the crossover procedure, which is obtained from  lattice QCD data. This figure tell us by passing the time the Universe becomes cooler  and the QGP is occurred (finished) at bout $0.2 -0.3$ nanosecond after the big bang.

\subsection{Low temperature region}
As mentioned above, the hadronic resonance gas (HRG) model yields a realistic equation of state in the low temperature regime, $T \lesssim 180$ MeV \cite{78}.  In the
 HRG scenario, the confinement phase of QCD  is treated
as a non-interacting gas of fermions and bosons \cite{79,80}.  The idea of the HRG model is to
implicitly account for the strong interaction in the confinement phase by looking at the hadronic
resonances only, since these are basically the only relevant degrees of freedom in that phase.
 The HRG result can be parameterized for the trace anomaly \cite{78}
\be\l{34}
{I(T) \ov T^4}\equiv  {\R -3p \ov T^4} = a_1T + a_2T^3 + a_3T^4 + a_4T^{10},
\ee
where $a_1$ = 4.654 GeV$^{-1}$, $a_2$ = -879 GeV$^{-3}$, $a_3$ = 8081 GeV$^{-4}$, $a_4$ = -7039000 GeV$^{-10}$.
In lattice QCD, through the computation of the trace anomaly $I(T) = \R(T) - 3p(T)$, one can estimate the pressure, energy density, and entropy density, with the help of the usual thermodynamic identities. The pressure difference at two temperatures $T$ and $T_{\rm low}$ can be expressed as the integral of the trace anomaly
\be\l{35}
{p(T) \ov T^4 } - {p(T_{\rm low})\ov T^4_{low}} = \int^T_{T_{\rm low}} {dT' \ov T'^5}I(T').
\ee
For sufficiently small values of the lower integration limit, $p(T_{\rm low})$ can be neglected due to the exponential
suppression \cite{81}. The energy density $\R(T) = I(T) + 3p(T)$  can then be calculated.  Using
Eqs.\;(\ref{34}) and (\ref{35}) we obtain
\be\l{36}
\R(T)= 3a_0T^4 + 4a_1T^5 + 2a_2T^7 + {7a_3 \ov 4}T^8 + {13a_4 \ov 10} T^{14},
\ee
\be\l{37}
p(T) =a_0T^4 + a_1T^5 + {a_2 \ov 3}T^7 + {a_3 \ov 4}T^8 + {a_4 \ov 10} T^{14},
\ee
where $a_0= -0.112$. In this case we  consider the era before the phase transition (quark-gluon phase) at low temperatures, in which the Universe is in the confinement phase and can be treated as a non-interacting gas of fermions and bosons \cite{79,80}. From the conservation Eq.(\ref{28})  we have
\begin{equation}\l{38}
H=- \frac{2\Big [12a_0 T^3+20a_1 T^4 + B_0(T)\Big]\dot{T}}{6\Big  [4a_0 T^4 + 5 a_1 T^5 + B_1(T)\Big] + m\xi B_2(T)},
\end{equation}
where
\begin{eqnarray}\l{39}
B_0(T)&=& 14a_2 T^6 + 14 a_3T^7 + \frac{91}{5} a_4 T^{13}, \\
B_1(T)&=& \frac{7}{3}a_2T^7 + 2a_3T^8 + \frac{7}{5}a_4T^{14},\l{40}\\
B_2(T)&=& a_1T^5 + a_2T^7 + a_3T^8 + a_4T^{14} + 4\lambda.\l{41}
\end{eqnarray}
To obtain the scale factor as a function of temperature we should integrate  Eq.(\ref{38}). To solve the Eq.\;(\ref{38}) we need   to make some approximating assumptions. Whereas the coefficient $a_2= -879 * 10^{-9} $MeV$^{-3}$, $a_3=8081* 10^{-12} $MeV$^{-4}$ and $a_4=-7039* 10^{-27} $MeV$^{-10}$, are very smaller than $a_0$ and $a_1$, then for obtaining the scale factor, we neglect those terms which the coefficients of them are  $a_2, a_3, a_4$. So  by this approximation, one can obtain the scale factor as

\begin{equation}\l{46}
a(T) \cong a_c U_1^{-{12a_0  /(30 + m\xi) a_1}} \exp\Bigg\{-U_0 {\rm Arctan}\Big[{(30+m\xi)a_1T +12a_0 \over \sqrt{m\xi(30+m\xi) a_1\lambda -36 a_0^2}}\Big]\Bigg\},
\end{equation}
where
\begin{eqnarray}
U_1&=&\Big [\big(5+\frac{m\xi}{6}\big)a_1T^2 +4 a_0T +\frac{2m\xi}{3}\lambda\Big], \\
U_0&=&\Big [{8(12+3m\xi)a_0\over (30+m\xi) \sqrt{m\xi(30+m\xi) a_1\lambda -36 a_0^2}}\Big].
\eea
Eqs.\;(\ref{26}), (\ref{36}), (\ref{37}) and (\ref{38}), the time derivative of temperature is
\begin{eqnarray}\l{51}
\dot{T}&=&\frac{- \bigg[6 \Big(4a_0 T^4 + 5 a_1 T^5 + B_1(T)\Big) - m\xi  B_2(T) \bigg] }{2 \sqrt{6\big (6-(N \xi a^\xi)^2\big)} \Big(12a_0 T^3+20a_1 T^4 + B_0(T) \Big)} \nonumber \\
 & \times &  \Bigg\{  \rho_m \Big(2\lambda + \rho_m \Big) + {3m^2\over 8\phi_0^2}\Big(3p-\rho-4\lambda\Big)^2    \Bigg\}^{\frac{1}{2}}.
\end{eqnarray}

This  expression describes the behavior of temperature  as a function of cosmic time in the QGP for the  brane gravity with chameleon scale field in the bulk.
In fact this relation shows the era before phase transition at low temperature where the universe is in
the confinement phase and is treated as a non-interacting gas of fermions and bosons.

\begin{figure}[ht]\label{1}
\centerline{ \includegraphics[width=7cm] {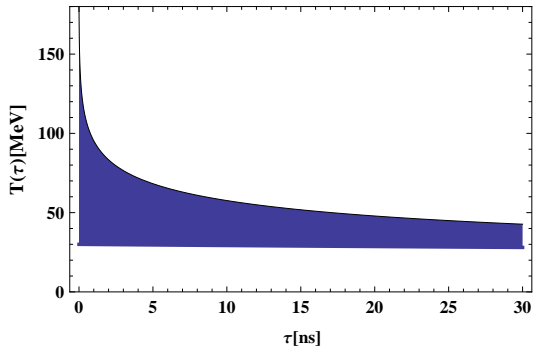}}
\caption{ $T$ versus $\tau$ according to the low temperature region of the  smooth crossover procedure in the brane gravity with chameleon scalar field in the bulk for $\lambda =10^9 $ MeV$^4$ in the low region.  We have set $N =0.2\times 10^3 $, $\xi = 0.002 $, and $m=1$.  }
\end{figure}
We   solved Eq.\;(\ref{51})  numerically and the result is plotted  in Fig.\;2 for $\lambda =10^9 $ MeV$^4$ in interval 30 MeV $ < T <$ 180 MeV. The plot shows two curve for Eq.(\ref{51}) related to two initial value of temperature as $T=180$ MeV and $T=30$ Mev, and the interval space between this two curves is shaded. This figure  shows the
behavior of the temperature as a function of cosmic time $\tau$. From this figure one can find that for low temperature region, $30\leq T\leq 180$, the QGP of the  Universe is occurred (finished) at about 15-30 nanosecond after the big bang. We see that QGP in the  low temperature  region in smooth crossover approach takes place after high temperature region, and this agrees with lattice QCD data analysis prediction.


\section{First order quark-hadron phase transition}
\ \ \ Whereas discretization effects in lattice calculations could be large, the HRG model in the intermediate temperature region 180 MeV $ \lesssim T \lesssim $250 MeV is no longer reliable and our study in the previous section shows this issue. Therefore the evolution of the Universe in the interval $180 \leq T < 250$  should to be studied  again using a non-crossover approach.

On the other hand, as  mentioned in Section \ref{III}, the quark-hadron phase transition in QCD is  characterized by the
singular behavior of the partition function, and may be the first or second order phase transition \cite{43}. In this section we assume that the quark-hadron phase transition is first order and occurs in the  intermediate temperature region $ T \lesssim 250$ MeV too.   Accordingly, we shall take  the  equation of state for matter in the quark-gluon phase to be given  by \cite{43}:
\begin{equation}\l{52}
\rho _q=3a_q T^4 +U(T), \ \ \ \ \ p_q =a_q T^4 -U(T).
\end{equation}
Here, the subscript $q$ denotes quark-gluon matter and   $a_q=61.75(\pi^2/90)$. The potential energy density $U(T)$  is  \cite{43}:
\begin{equation}\l{53}
U(T)=B+\gamma_T T^2 - \alpha_T T^4,
\end{equation}
where $B$ is the bag pressure   constant, $\alpha_T = 7\pi^2/20$, and
$\gamma_T = m^2_s/4$, where $m_s$, the mass  of the strange quark,  is in the
$60 - 200$ MeV range. This form of  $U$ comes from a model in which the quark fields
interact with a chiral field formed by  the $\pi$ meson field together with a scalar field.
Results obtained from low energy hadron spectroscopy, heavy ion collisions, and  from phenomenological fits of light hadron
properties give a value of $B^{1/4}$ between 100 and 200 MeV \cite{24}.

In the hadron phase, one takes the cosmological fluid to be   an ideal gas of massless pions and  nucleons
described by the Maxwell-Boltzmann  distribution function with energy density $\rho_h$ and pressure $p_h$. Hence, the equation of state in the hadron phase is
\begin{equation}\l{54}
 p_h ={1 \over 3}\rho _h=a_{\pi} T^4,
\end{equation}
where $a_{\pi} =17.25 (\pi^2/90)$.

The critical temperature $T_c$ is defined by the condition $p_q (T_c) = p_h (T_c)$ \cite{76}. Taking   $m_s = B^{1/4} = 200$ MeV, the critical temperature  is
\begin{equation}\l{55}
 T_c={\Bigg[{\frac{\gamma_T + \sqrt{\gamma^2_T+ 4B(a_q+\alpha_T-a_{\pi})}}{2(a_q+\alpha_T -a_{\pi})}}\Bigg]}^{1\over 2}\approx 125\   {\rm MeV}.
\end{equation}
 Since the phase transition is  first order, all physical quantities, such as  the energy density, pressure, and entropy, exhibit discontinuities across the critical curve.


\subsection{Behavior of temperature in the QGP for general $U(T)$ }
In order to investigate the quark-hadron phase transition in this section, we shall first obtain the scale factor by using the conservation relation and an especial equation of state,  as in the  previous section. Using Eqs.\;(\ref{52}), (\ref{53}),  and  the conservation relation (\ref{28}), we get
\begin{equation}\l{56}
H=- \frac{\Big[ 2(3a_q-\alpha_T)T^3+\gamma_T T \Big] \dot{T}}{(6a_q -m\xi \alpha_T)T^4+m \xi   \gamma_T T^2 +m\xi ( B + \lambda) }.
\end{equation}
By integrating Eq.\;(\ref{56}), the scale factor can be acquired as a function of temperature. Namely
\begin{eqnarray}\l{57}
a(T)&=&a_c \Big[ U0 T^4 + m\xi\gamma_T T^2 + m\xi(B+\lambda)\Big]^{a_T-3a_q\over 2U_0}\\&&\times
\exp\bigg\{ {U_1\over 2U_0U_2}{ \rm Arctan}\Big[{ 2U_0T^2 +m\xi \gamma_T\over U_2}\Big]\bigg\},\nonumber
\end{eqnarray}
where,
\begin{eqnarray}
U_0 &=& 4\Big(3a_q-\alpha_T\Big), \\
U_1&=& -6 \gamma_T a_q\Big(2-m\xi\Big),\\
U_2&=& \Big[4m\xi(B+\lambda)(6a_q-m\xi \alpha_T) -m^2\xi^2\gamma^2_T \Big]^{1/2}.
\end{eqnarray}
Therefore, the Freidmann equation and the conservation relation lead us to following relation for $\dot{T}$
\begin{eqnarray}\l{58}
\dot{T}&=&-\frac{ (6 a_q - m\xi\alpha_T)T^4 + m\xi \gamma_T T^2 + m\xi(B - \lambda) }{\sqrt{6(6-(\xi\phi_0)^2)}\Big[ 2(3a_q-\alpha_T)T^3+\gamma_T T \Big]} \\
 & & \qquad \times\Bigg\{   \rho_q(T) \Big(2\lambda+ \rho_q(T) \Big) + \lambda\Big( \frac{m}{\phi} \Big)^2 (\alpha_T T^4 - \gamma_T T^2 -B-\lambda)^2 \Bigg\}^{\frac{1}{2}}  \nonumber
\end{eqnarray}
where $\rho_m(T)=(3a_q-\alpha_T)T^4 + \gamma_T T^2 + B$. We have plotted the numerical results of Eq.\;(\ref{58}) in Fig.\;3 for $\lambda=10^9  $ MeV$^4$(solid line), $\lambda=5\times 10^9  $ MeV$^4$(Dashed), $\lambda= 10\times10^9  $ MeV$^4$(Dotted line), $\lambda=15\times 10^9  $ MeV$^4$(Dashed-dotted line). This figure shows  that the effective temperature of the Universe in the quark-gluon phase (QGP) decreases with passing time and reaches to the critical temperature   at about (0.05-  0.2) nanosecond  after the big bang. Moreover by increasing the brane tension the  decreasing of the   temperature  will be  faster.

\begin{figure}[ht]\label{3}
\centerline{ \includegraphics[width=7cm] {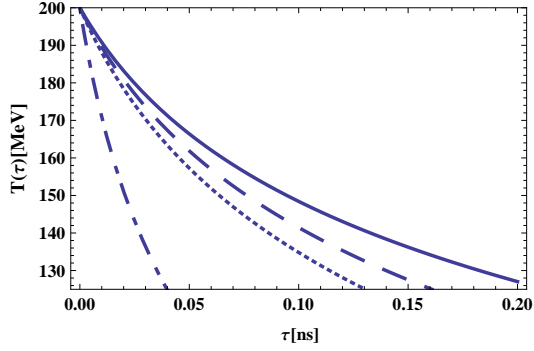}}
\caption{   $T$ versus $\tau$ in the quark-gluon  phase for $\lambda=10^9  $ MeV$^4$(solid line), $\lambda=5\times 10^9  $ MeV$^4$(Dashed), $\lambda= 10\times10^9  $ MeV$^4$(Dotted line), $\lambda=15\times 10^9  $ MeV$^4$(Dashed-dotted line) in the brane world gravity with chameleon scalar field in the bulk.  We have set $N =0.2\times 10^4 $, $\xi = 0.002 $, and m=1. }
\end{figure}

\subsection{Behavior of temperature in the QGP for  $U(T)=B$ }
One popular model that deals with quark confinement,  is that of an elastic bag which allows the quarks to move  around freely, and the potential energy density is constant. In this case, the    equation of state  for the  quark matter is given by the bag model, namely  $p_q = {(\rho_q -4B) / 3}.$   Therefore, using  the conservation relation, we get that
\begin{equation}\l{59}
H=- \frac{6a_q T^3 \dot{T}}{6a_qT^4+C_1 },
\end{equation}
in which $$C_1=m\xi \Big(B+\lambda\Big).$$
This first order differential equation provides us with the  scale factor function
\begin{equation}\l{60}
a(T)=a_c \Big( 6a_q T^4 + C_1 \Big)^{-{1/4}}
\end{equation}
Finally, using the Friedmann equation, we find that the time evolution of the temperature $T$ is described as follows
\begin{eqnarray}\l{61}
\dot{T}&=&-\frac{ (6a_qT^4+C_1)}{6a_q T^3 \sqrt{6\Big(6-( \xi \phi_0)^2\Big)}}  \\
   & \times& \Bigg\{ \Big(3a_qT^4 + B\Big) \Big(2\lambda+3a_qT^4 + B \Big)    +  6 \frac{m^2}{\phi^2} \Big(B+\lambda\Big)^2
    \Bigg\}^{1/2} \nonumber
  \end{eqnarray}
\begin{figure}[ht]\label{4}
\centerline{ \includegraphics[width=7cm] {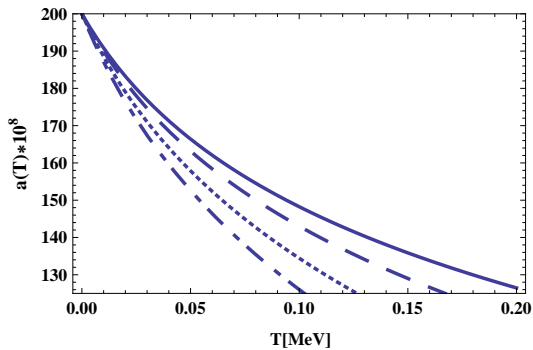}}
\caption{  $T$ versus $\tau$ in the quark-gluon  phase for $\lambda=10^9  $ MeV$^4$(solid line), $\lambda=5\times 10^9  $ MeV$^4$(Dashed), $\lambda= 10\times10^9  $ MeV$^4$(Dotted line), $\lambda=15\times 10^9  $ MeV$^4$(Dashed-dotted line)in the brane world gravity with chameleon scalar field in the bulk.  We have set $N =0.2\times 10^4 $, $\xi = 0.0002 $, m=1,  and $U(T)=B$.  }
\end{figure}
We have plotted the numerical results of Eq.\;(\ref{61}) in Fig.\;4 for $\lambda=10^9  $ MeV$^4$(solid line), $\lambda=5\times 10^9  $ MeV$^4$(Dashed), $\lambda= 10\times10^9  $ MeV$^4$(Dotted line), $\lambda=15\times 10^9  $ MeV$^4$(Dashed-dotted line). In this case, $U(T) = B$, the decreasing of temperature is slower than the general case of $U(T)$. This figure shows  that the effective temperature of the Universe decreases with passing time in the QGP at about (0.1-0.2) nanosecond after the big bang. One can see that the QGP in the high temperature region in the smooth crossover approach is very similar with the QGP in the first order phase transition formalism but QGP in the low temperature region of smooth crossover approach has taken place later with respect to high temperature region of crossover approach and first order phase transition formalism.

\subsection{Behavior of the hadron volume fraction}
During the phase transition, the temperature, pressure, enthalpy and entropy  are conserved and the  quark matter density decreases from $\rho_Q$ to $\rho_H$, the hadron matter density. During this transition, we have
$T_c=125$ MeV, $\rho_Q = 5 \times 10^9 $ MeV$^4$, $\rho_H \approx 1.38 \times 10^9 $ MeV$^4$, and constant pressure
$p_c \approx 4.6 \times 10^8 $ MeV$^4$.

In this step, the energy density is described by the volume fraction of matter, namely
\begin{equation}\l{62}
\rho(\tau) = \rho_H h(\tau)+\rho_Q \Big(1-h(\tau)\Big).
\end{equation}
At the beginning of the phase transition, we have $\rho(\tau_c)=\rho_Q$, $h(\tau)=0$,  and the whole of the matter in the Universe is described by quarks. However, at the end of the phase transition, the whole of matter is described by hadrons $\rho_H$, namely $\rho(\tau_h)=\rho_H$, $h(\tau_h)=1$, and the Universe enters its hadron phase.

The conservation relation implies that
\begin{equation}\l{63}
H=-\frac{r\dot{h}}{D_0 rh + D_1},
\end{equation}
where:
$$D_0=3\Big(1+\frac{m}{6}\xi\Big)$$ $$D_1=3\Bigg[ 1-\frac{m}{6} \xi \Big( \frac{3p_c-\rho_Q-4\lambda}{\rho_Q+p_c} \Big) \Bigg],$$
and $r$ is defined as
$$r=\frac{\rho_H-\rho_Q}{p_c+\rho_Q}.$$
The scale factor is easily obtained as a function of hadron volume fraction
\begin{equation}\l{64}
a(T)= a_c \Big( D_0rh+D_1 \Big)^{-{1/D_0}}.
\end{equation}
We find out that the Friedmann equation produces  the evolution of the hadron fraction during the quark-hadron phase transition
\begin{eqnarray}\l{65}
\dot{h}&=& -\frac{D_0 rh + D_1}{r\sqrt{6(6-(N \xi a^\xi)^2)}} \Bigg\{  \Big[\rho_Q+(\rho_H-\rho_Q)h\Big]\Big( 2\lambda + \rho_Q+(\rho_H-\rho_Q)h \Big)  \nonumber \\
 & & \qquad \qquad \qquad \qquad \qquad + 6\Big( \frac{m}{4\phi} \Big)^2 \Big(3p_c-\rho_Q-(\rho_H-\rho_Q)h-4\lambda\Big)^2 \Bigg\}^{1/2}.
\end{eqnarray}
\begin{figure}[ht]\label{5}
\centerline{ \includegraphics[width=7cm] {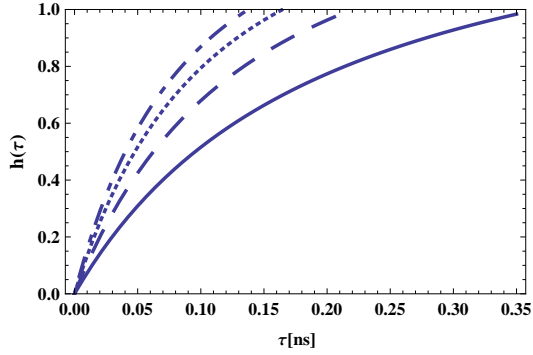}}
\caption{  Hadron volume fraction as a function of  cosmic time $\tau$ for $\lambda=10^9  $ MeV$^4$(solid line), $\lambda=5\times 10^9  $ MeV$^4$(Dashed), $\lambda= 10\times10^9  $ MeV$^4$(Dotted line), $\lambda=15\times 10^9  $ MeV$^4$(Dashed-dotted line)in the brane world gravity with chameleon scalar field in the bulk.  We have set $N =0.2\times 10^4 $, $\xi = 0.0002 $, and m=1. }
\end{figure}
We presented  the numerical results of Eq.\;(\ref{65}) in Fig.\;5 for$\lambda=10^9  $ MeV$^4$(solid line), $\lambda=5\times 10^9  $ MeV$^4$(Dashed), $\lambda= 10\times10^9  $ MeV$^4$(Dotted line), $\lambda=15\times 10^9  $ MeV$^4$(Dashed-dotted line). This figure shows  that the hadron volume fraction  increases with passing  time during the quark-hadron phase transition QHPT and it take about 0.1-0.35 nanosecond. This figure completely agrees with the results of previous figures and by increasing the brane tension the increasing of the hadron volume fraction  will be faster. We plot the scale factor of the Universe during the QHPT as a function of hadron volume fraction, $h$, in Fig.\;6. From figure\;6  is seen that during the QHPT  the Universe is expanding, although temperature, pressure, enthalpy, and entropy  during the phase transition are constant.

\begin{figure}[ht]\label{6}
\centerline{ \includegraphics[width=7cm] {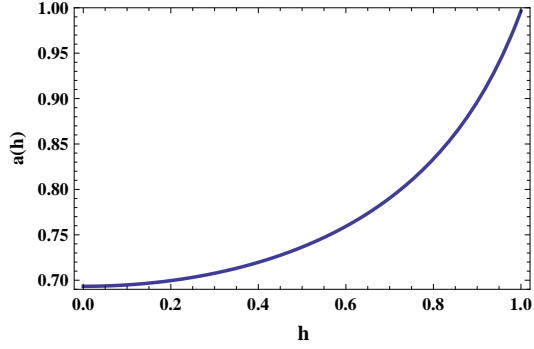}}
\caption{  Scale factor as a function of the hadron volume fraction  during the QHPT in the brane world gravity with chameleon scalar field in the bulk.   We have set $N =0.2\times 10^4 $, $\xi = 0.0002 $, and m=1. }
\end{figure}

\subsection{Behavior of temperature in the hadron phase}
In this stage of its evolution, the Universe is in a  pure hadronic phase, with an equation of state given by
\begin{equation}\l{66}
\rho=3p=3a_\pi T^4.
\end{equation}
Substituting this equation of state into the  conservation relation leads to the following differential equation
\begin{equation}\l{67}
H= -\frac{6a_\pi T^3 \dot{T}}{6a_\pi T^4 +m\xi \lambda }.
\end{equation}\l{68}
 By integrating  Eq. (\ref{65}), one obtains the scale factor as a function of temperature
\begin{equation}\l{69}
a(T)= a_c \Big( 6 a_\pi T^4+m\xi\lambda \Big)^{-{1/4}}.
\end{equation}
Consequently, we get that $\dot{T}$, in the pure hadronic phase era,  is
\begin{eqnarray}\l{70}
\dot{T} = -\frac{6a_\pi T^4+2m\xi\lambda}{6a_\pi T^3 \sqrt{6\big[6-(N \xi a^\xi)^2\big]} }  \Bigg\{3 a_{\pi} T^4 \Big( 2\lambda+ 3a_{\pi}T^4 \Big) +{6 m^2\over \phi_0^2} \Bigg\}^{1/2}
\end{eqnarray}
\begin{figure}[ht]\label{0}
\centerline{ \includegraphics[width=7cm] {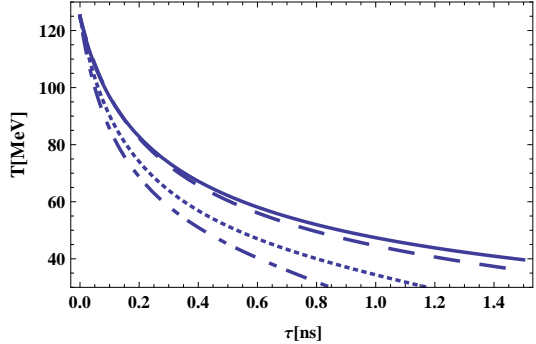}}
\caption{  $T$ versus $\tau$ in the pure hadronic phase era for  $\lambda=10^9  $ MeV$^4$(solid line), $\lambda=5\times 10^9  $ MeV$^4$(Dashed), $\lambda= 10\times10^9  $ MeV$^4$(Dotted line), $\lambda=15\times 10^9  $ MeV$^4$(Dashed-dotted line)in the brane world gravity with chameleon scalar field in the bulk.  We have set $N =0.2\times 10^4 $, $\xi = 0.0002 $, m=1. }
\end{figure}
We plotted the numerical results of Eq.\;(\ref{70}) in Fig.\;7 for $\lambda=10^9  $ MeV$^4$(solid line), $\lambda=5\times 10^9  $ MeV$^4$(Dashed), $\lambda= 10\times10^9  $ MeV$^4$(Dotted line), $\lambda=15\times 10^9  $ MeV$^4$(Dashed-dotted line).  This figure shows  the effective temperature of the Universe  as a function of cosmic time in the brane gravity with chameleon scalar field in the bulk for hadron area. This figure shows that the hadron area is occurred at about 0.8-1.5   nanosecond after the big bang and this fact    confirms the results of  other areas.\\
\section{Discussion and Conslusion}
\ \ \ In this paper we considered the quark-hadron phase transition in brane-world model including a bulk scalar field. The peroblem of quark-hadron phase transition has been investigated in different models of brane-world scenario.
Quark-hadron phase transition in RS brane-world model has been performed in \cite{24}. It was shown that the temperature of the early universe in brane-world model in smaller than standard model, and small value of brane tension reduces the quark-gluon plasma temperature and accelerate  phase transition to hadron era. In addition, hadron fraction, $h$, strongly depends on brane tension and small value of brane tension leads to bigger value of $h$ relative to standard model. In \cite{24} the effect of dark radiation also has been considered and it was obtained that large value of dark radiation strongly accelerate the formation of hadron phase. In \cite{65}, a Brans-Dicke(BD) brane-world model has been utilized to investigate quark-hadron phase transition. The temperature of the early universe is smaller than RS brane-world model. Quark-gluon plasma temperature is reduced by increasing the BD coupling constant, $\omega$. The hadron fraction depends on the BD coupling as well in which large value of coupling constant leads to higher $h$ and finally smaller time interval is necessary for phase transition. {\bf We} also considered variation of temperature by using crossover approach in high and low temperature. In smooth crossover regime, where lattice QCD for high temperature was used, there is a smooth slope relative to first order phase transition and for low temperature where HRG is used, the slope is steep relative to first order phase transition.\\
The framework which has been utilized in our work is a generalized form of \cite{65} in which the bulk scalar field has a nan-minimal coupling to matter as well as a non-minimal coupling to geometry.
Many authors have tried to add some extra terms to the energy momentum for obtaining a brane world model with the bulk-brane energy transfer, but in our model  the energy transfer is caused by interaction between the  scalar field in the bulk and  the matter  field on the brane. In fact, due to this bulk-brane energy transfer and modified  Friedmann equation the rate of expansion of the Universe  is increased  in the early time. In this work we studied the  QCD phase transition with two different mechanisms; smooth crossover approach and first-order phase transition formalism. We took into  consideration  the dynamical evolution of physical quantities such as, energy density, effective temperature and scale factor,before,  during and after the phase transition. Our analysis shows that, the quark-hadron phase transition has taken place at about nanosecond after the big bang. First we studied the QCD phase transition utilizing smooth crossover formalism in two regime of high and low temperature. Equation of state is an important relation in quark hadron phase transition. A lattice data can be utilized for trace anomaly to construct an appropriate EoS. In high temperature regime, namely $T \geq 250$MeV the trace anomaly can be computed accurately which shows a radiation like behavior; however in low temperature regime $T \leq 150$MeV the trance anomaly is affected by large discretization effect. So HRG model is used to build a realistic EoS in this regime. Considering the results in detailed indicate some differences. Our analysis shows that in  the high temperature region of the smooth crossover  the effective temperature of the  QGP of the Universe is  in the interval MeV $250 \leq T \leq 700$ MeV   and QGP has occurred (finished) at about 0.2-0.3 nanosecond after the big bang. And in the low temperature region of the smooth crossover  the effective temperature of the  QGP of the Universe is  in the interval MeV $50 \leq T \leq 180$ MeV and QGP has occurred (finished)  at about 15 nanosecond after the big bang. Note that this result is about $10^{-3}$ times earlier than the prediction of other works \cite{58, 60}.\\
Then we studied phase transition using the first order phase transition formalism where we have three stages; before (QGP), during and after (hadron) phase transition. We find out that for two different models; general $U(T)$ and bag model with $U(T)= constant$, the effective temperature of the QGP is in the interval MeV $125 \leq T \leq 200$ MeV  at about  0.05-0.2  nanosecond after the big bang. It is notable that the QGP in the  bag model took place a little faster than general $U(T)$.  Furthermore our analysis shows that the phase transition in the first order phase transition formalism took about 0.35 nanosecond and during the quark-hadron phase transition (QHPT) the Universe is expanded although the temperature, pressure, enthalpy and entropy of the system are constant. Finally we examine that the hadronic area takes place in the interval  MeV $30 \leq T \leq 125$ MeV at about 0.8-1.4 nanosecond  after the big bang. The results of this case was solve numerically and depicted for different values of brane tension. It was realized that large value of brane tension significantly reduce the temperature of early universe and accelerate the phase transition. The relation between temperature and brane tension in this work is unlike \cite{24}. It seems that the mechanism which brung such a result is the coupling of scalar field to other companents. At last our study shows that the general behavior of the  effective temperature of the Universe is similar in the first order phase transition formalism and smooth crossover approach.

\section{Acknowledgement}
The authors would like to thank Shawn Westmoreland for helping in writing  the paper in good English.


\end{document}